\documentstyle[11pt,fleqn]{article}
\def\ref{par\noindent\hangindent=6mm\hangafter=1}
\begin{document}
\baselineskip 8mm
%
\begin{center}
{\scriptsize Nuovo Cimento B 113 (1998) 119-124 \& gr-qc/9611039}\\
{\bf WHITTAKER QUANTUM UNIVERSES}

\bigskip

H. Rosu\footnote{E-mail:
rosu@ifug3.ugto.mx; Fax: 0052-47-187611}
and J. Socorro\footnote{E-mail:
socorro@ifug4.ugto.mx}

{\it Instituto de F\'{\i}sica de la Universidad de Guanajuato, Apdo Postal
E-143, Le\'on, Gto, M\'exico}


\end{center}

\bigskip


{\bf Summary.} - We show that closed, radiation-filled
Friedmann-Robertson-Walker quantum universes of arbitrary factor ordering
obey the Whittaker equation. We also present the formal Witten
factorization as well as the double Darboux strictly isospectral scheme for
the Whittaker equation.



\vskip 2cm


If in the ``time-time" component of Einstein's equations
for Friedmann-Robertson-Walker (FRW) universes
$\dot a^{2}+k=\frac{8\pi G}{3}\rho a^2$ one substitutes a canonical
momentum $\pi _{a}$ conjugate to the scale factor and quantizes the system
according to the common procedure
$\pi _{a}\rightarrow -i(d/d a)$, one
will obtain the corresponding Wheeler-DeWitt (WDW) equation

$$
\Bigg[-a^{-p}\frac{d}{d a}a^p\frac{d}{d a} +
\left(\frac{3\pi}{2G}\right)^2\frac{1}{k^3}\left(ka^2-\frac{8\pi G}{3}
\rho a^4\right)
\Bigg]u(a)=0~.
\eqno(1)
$$

The parameter $p$ enters as a consequence of the ambiguity in the ordering
of $a$ and $d/d a$. Amongst the most used orderings,
the $p=1$ is the so-called Laplacian one, which has been used for the
two-dimensional mini-superspace describing a massive scalar field in a closed
universe, i.e., a model obtained through the substitution
$\frac{8\pi G}{3}\rho\rightarrow m^2\phi ^2$, $k=1$ in Eq. (1) and adding
$\frac{1}{a^2}\frac{\partial ^2}{\partial \phi ^2}$ to the $a$ derivatives
turned into partial ones too, while the dependent variable
becomes $u(a,\phi)$ \cite{haw1}.
On the other hand, the $p=2$ ordering is convenient for
the semiclassical approximation \cite{hp}.

In the matter sector various contributions may be introduced but here we
shall consider only the radiation case, i.e.,

$$
\frac{8\pi G}{3}\rho\rightarrow \frac{8\pi G}{3}\Bigg[\rho _r
\left(\frac{a_0}{a}\right)^4\Bigg]~,
\eqno(2)
$$
where $\rho _r$ is the radiation energy density.
The most compact form of the WDW equation in the FRW metric can be written
down as follows

$$
\Bigg[-\tilde{a}^{-p}\frac{d}{d \tilde{a}}\tilde{a} ^p
\frac{d}{d \tilde{a}} + V(\tilde{a}) \Bigg] u (\tilde{a})=0~,
\eqno(3)
$$
where the tilde variables are rescaled and dimensionless ones.
Thus, only with the radiation matter sector
the WDW ``potential'' function turns out to be of the form

$$
V(\tilde{a})\equiv\tilde{a} ^2   -
\tilde{\beta}^2~,
\eqno(4)
$$
where $\tilde{a} ^2=\frac{3\pi}{2Gk}a^2$ is the tilde scale factor of the
universe, and $\tilde{\beta} ^2=\frac{4\pi ^2}{k^2}\rho _ra_0^4$ expresses
the radiation effect on the cosmological expansion; $k$ is considered
positive (closed universes).

With the ansatz $u(\tilde{a})\equiv g(\tilde{a})e^{-\tilde{a}^2/2}$, and
with the further change of variable $x=\tilde{a}^2$, one gets the
confluent hypergeometric equation for $g(\tilde{a})$ as follows

$$
x\frac{d^2g}{dx^2}+\left(\frac{1+p}{2}-x\right)\frac{dg}{dx}+
\left(\frac{\tilde{\beta}^2-1-p}{4}\right)g=0~.
\eqno(5)
$$
The complete solution of (5) is the superposition of
confluent hypergeometric (Kummer) functions $M$ and $U$ \cite{as}
$g(\tilde{a})=C_{m}M(\frac{1+p-\tilde{\beta}^2}{4},\frac{1+p}{2};\tilde{a})
+C_{u}U(\frac{1+p-\tilde{\beta}^2}{4},\frac{1+p}{2};\tilde{a})$, where $C_{m}$
and $C_{u}$ are superposition constants.
Thus, mathematically speaking, the problem may be considered as solved up to
fixing some boundary conditions.
However, for several particular purposes (such as Darboux constructions,
see below)
one should make another mathematical step, which is to pass to a
self-adjoint form of the confluent equation.
This has first been done
by Whittaker in 1904 \cite{wh} by eliminating the first derivative term,
i.e., by substituting
$g=|\tilde{a}|^{-\frac{p+1}{4}}\exp(-\tilde{a}/2)y(\tilde{a})$
in the confluent equation (one can also get a self-adjoint form by
multiplying the confluent
equation by the weight function $\tilde{a} ^{\frac{p-1}{2}}e^{-\tilde{a}}$).
The Whittaker equation reads
$$
y_{\lambda , \mu}^{''}(\tilde{a})-\frac{1}{4}
\left(1-\frac{\tilde{\beta}^2}{\tilde{a}}-
\frac{1-[(p-1)/2]^2}{\tilde{a}^2}\right)y_{\lambda , \mu}(\tilde{a})=0~,
\eqno(6)
$$
where
$\lambda=(\tilde{\beta}/2)^2$, $\mu=(p-1)/4$,
and the general solution is the linear combination
$y_{\lambda , \mu}=C_1 M_{\lambda ,\mu}+C_2W_{\lambda ,\mu}$ \cite{as}.
Notice that
$V_{confl}=\frac{1}{4}\left(1-\frac{\tilde{\beta}^2}{\tilde{a}}-
\frac{1-[(p-1)/2]^2}{\tilde{a}^2}\right)$ can be thought of as a
Schr\"odinger potential, and the Whittaker equation as a Schr\"odinger
equation at zero energy.
The two Whittaker functions are defined as follows \cite{wh}
$$
M_{\lambda , \mu}(\tilde{a})=
e^{-\tilde{a}/2}\tilde{a}^{\mu+1/2}M(\mu -\lambda +\frac{1}{2},
2\mu+1;\tilde{a})
\eqno(7)
$$
and
$$
W_{\lambda ,\mu}(\tilde{a})=\frac{\Gamma (2\mu)}{\Gamma(\mu -\lambda +1/2)}
M_{\lambda , -\mu}(\tilde{a})+
\frac{\Gamma (-2\mu)}{\Gamma(-\mu -\lambda +1/2)}
M_{\lambda ,\mu}(\tilde{a})~,
\eqno(8)
$$
where $M$ is the Kummer function. In Eq. (8) the right-hand side is replaced
by its limiting value when $2\mu$ is an integer or zero.
As is well-known the problem of boundary conditions in quantum cosmology is
more involved than in quantum mechanics. The most used choices are the
Hartle-Hawking ``no boundary" one \cite{hh} and
Vilenkin's ``tunneling" condition \cite{v}.
At the level of the Whittaker equation the choice of the boundary conditions
depends on what feature one would like to emphasize more. For example, in
atomic physics the physically acceptable solution is the $W_{\lambda ,\mu}$
function only because it is exponentially declining, though singular at the
origin. However, in quantum cosmology the exponentially growing
$M_{\lambda ,\mu}$ is also an acceptable solution. Suppose we would like
to adopt Lema\^{\i}tre's Primeval Atom paradigm \cite{le}. In that case one
selects $W_{\lambda ,\mu}$ and gives ``atomic" meaning to the subscripts,
i.e., $\lambda$ should be considered as an effective principal quantum
number and $\mu =l+1/2$, where $l$ stands for an angular quantum number.
As a matter of fact, the Rydberg states of the Lema\^{\i}tre atom are
the most interesting for the physical case. For this, one should use the
asymptotic properties of $W_{\lambda ,\mu}$ at large $\lambda$ and
$\mu$ \cite{olv}.

Let us address the problem of particular cases.
Since both Whittaker functions can be expressed in terms of the Kummer
function $M$, there is nothing new for this issue with respect to the
confluent equation. In particular for polynomial solutions the constraints
are $p\neq -2m-1$ and the ``quantization" conditions
$\tilde{\beta}^2=4n+p+1$, where $n$ is the
polynomial degree \cite{as}.
For example, the $p=1$ factor ordering leads to Laguerre polynomials
whereas the p=2 factor-ordering provides Hermite polynomials \cite{K}.
Many other special cases are given in \cite{as}.

The Whittaker form of the confluent equation is needed in order to
perform the single Darboux (Witten \cite{w}) construction and the
double Darboux one \cite{m}. However, the single Darboux construction
might be useless
from the physical point of view if the Witten superpotentials are
singular. This would also mean singular partner potentials. To see this,
suppose we would like to use $M_{\lambda ,\mu}$ for
the single Darboux construction, which is nothing but a factorization of the
Whittaker equation \cite{Vil}. The superpotential is the negative of the
logarithmic derivative of the $M_{\lambda ,\mu}$. This can be shown to be
$\frac{1}{2}-\frac{\mu +\frac{1}{2}}{\tilde{a}}-\frac{a}{b}
\frac{M(a+1,b+1;\tilde{a})}{M(a,b;\tilde{a})}$. One can find zones in the
$(a,b)$ -plane where $M(a,b;\tilde{a})$ has no zeros (see Fig. 13.1 in
\cite{as}). However,
the singularity in the first term at the branch point $\tilde{a}=0$
cannot be deleted unless $\mu =-1/2$ for
which the function $M_{\lambda , \mu}$ is actually undefined
(one can also see that the second confluent parameter should be nought,
$b=0$, and therefore the ratio $a/b$ becomes singular). For all the cases
$2\mu =\pm 1,\pm 2,...$
one must work with the function $W_{\lambda ,\mu}$, but again one should know
the nodeless regions in the parameter plane.
Despite this difficulty, the Witten factorization of
the Whittaker equation can be performed formally as an intermediate step
for the double Darboux transform. It is accomplished by means of the operators
$A_{1}=\frac{d}{d\tilde{a}}+{\cal W}(\tilde{a})$ and
$A_{2}=-\frac{d}{d\tilde{a}}+{\cal W}(\tilde{a})$, where
${\cal W}$ denotes the superpotential function.
The initial Riccati
equation of the Witten scheme is $V_{confl}={\cal W}^2- {\cal W}^{'}$ and
the partner (``fermionic") equation is $V_{f, confl}=
{\cal W}^{2}+{\cal W}^{'}$, corresponding to the ``fermionic" Whittaker
equation for which the factoring operators are applied in reversed order.
On the other hand, the double Darboux construction
can be performed with the general
Whittaker solution $y_{\lambda , \mu}$ without any constraints,
because the singularities introduced by the solutions
with zeros do not appear in the final result \cite{suk}.
The one-parameter family of strictly isospectral Whittaker potentials will be
$$
V_{iso}(\tilde{a};\epsilon)=\frac{1}{4}
\left(1-\frac{\tilde{\beta}^2}{\tilde{a}}-
\frac{1-[(p-1)/2]^2}{\tilde{a}^2}\right)
-\frac{4y_{\lambda ,\mu}y_{\lambda ,\mu}^{'}}
{{\cal J} _{\lambda ,\mu}+\epsilon}+
\frac{2y_{\lambda , \mu}^4}{({\cal J} _{\lambda ,\mu}+\epsilon)^2},
\eqno(9)
$$
where ${\cal J} _{\lambda ,\mu}(\tilde{a})\equiv\int_0^{\tilde{a}}
y_{\lambda ,\mu}^2(z)dz$
and $\epsilon$ is the family parameter, which is a real, positive number.
The derivative $y_{\lambda ,\mu} ^{'}$ can be written down as follows
$$
y_{\lambda ,\mu} ^{'}=\frac{2\tilde{a}-\tilde{\beta}
 ^2}{4\tilde{a}}y_{\lambda ,\mu}
-\frac{1}{\tilde{a}}y_{\lambda +1,\mu}+
C_{1}\frac{5+p+\tilde{\beta} ^2}{4\tilde{a}}M_{\lambda +1,\mu}~,
\eqno(10)
$$
where $C_1$ is the first superposition constant in the general Whittaker
solution.
For the double Darboux construction in the particular case
$p=1$ see \cite{rs}.

The wavefunctions of the Whittaker stictly isospectral one-parameter modes
can be written down as follows \cite{suk,rs}
$$
y_{\lambda , \mu, iso}\propto
\frac{y_{\lambda ,\mu}}{{\cal J} _{\lambda ,\mu}+\epsilon}~.
\eqno(11)
$$

We can hint on the following physical picture. Since the strictly
isospectral scheme introduces in principle an infinity of Whittaker modes
of the common WDW zero energy cosmology, one may try to identify those
modes with a class of ``radiation-only-containing" cosmic structures.
These structures are characterized by the mode numbers $\lambda$ and $\mu$,
i.e., by their radiation content and quantum factor ordering, by the set of
superposition constants, and by the
isospectral family parameter, which is a sort of decoherence parameter.
Moreover, following techniques presented in \cite{suk}, one can introduce
multiple-parameter families of strictly isospectral modes, that is more
decoherence parameters.


Finally, we remark that the class of Whittaker-WDW equations holds not
only for radiation-filled universes. Some other matter sectors can be
introduced without changing the confluent character of the equation.
However, the radiation-filled case is well suited for
resonator-cavity models of the universe at ``quantum" gravitational scales
\cite{ros}.

\begin{center}    ***    \end{center}
This work was partially supported by the CONACyT Projects 4868-E9406
and 3898-E9608.






\end{document}